# OAR-Weighted Dice Score: A spatially aware, radiosensitivity aware metric for target structure contour quality assessment

Lucas McCullum[1,2], Kareem A. Wahid[2,3], Barbara Marquez[1,4], Clifton D. Fuller[2]

[1]UT MD Anderson Cancer Center UTHealth Houston Graduate School of Biomedical Sciences, Houston, USA
[2]Department of Radiation Oncology, The University of Texas MD Anderson Cancer Center, Houston, USA
[3]Department of Imaging Physics, The University of Texas MD Anderson Cancer Center, Houston, USA
[4]Department of Radiation Physics, The University of Texas MD Anderson Cancer Center, Houston, USA

**Abstract** The Dice Similarity Coefficient (DSC) is the current *de facto* standard to determine agreement between a reference segmentation and one generated by manual / auto-contouring approaches. This metric is useful for non-spatially important images; however, radiation therapy requires consideration of nearby Organs-at-Risk (OARs) and their radiosensitivity which are currently unaccounted for with the traditional DSC. In this work, we introduce the OAR-DSC which accounts for nearby OARs and their radiosensitivity when computing the DSC. We illustrate the importance of this through cases where two proposed contours have similar DSC, but lower OAR-DSC when one contour expands closer to the surrounding OARs. This work is important because the OAR-DSC may be used by deep learning auto-contouring algorithms in a radiation therapy specific loss function, thereby progressing on the current disregard for the importance of these differences on the final radiation dose plan generation, delivery, and risks of patient toxicity.

## 1 Introduction

Automated structure contouring, also known as auto-contouring, is a method to automatically extract desired regions in an image. Typically, ground truth contours are made manually by in-field experts and compared to the predictions from an auto-contouring algorithm. One standard *de facto* metric of measuring the agreement between the ground truth and predicted segmentations in the current era of deep learning is the Dice Score, or more formally the Dice Similarity Coefficient (DSC) [1,2], as described in Figure 1.

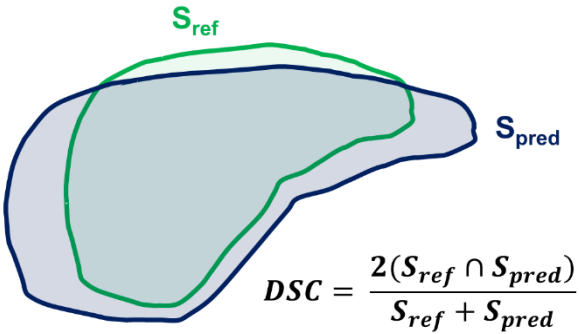

*Figure 1: Formulation of the DSC between the reference contour, $S_{ref}$, and the predicted auto-contour, $S_{pred}$.*

The DSC has subsequently extended to and now dominates the relevant literature in the field of radiation oncology to compare ground truth segmentations of tumors and surrounding normal tissue sensitive to unintended radiation, also known as Organs-at-Risk (OAR). Since deep learning models are often trained using a negative / positive feedback system where lower DSC is penalized and higher DSC is reinforced, equivalent DSC will produce the same feedback during model training, regardless of where the errors occurred. This is because the original formulation of DSC does not account for spatial information since it did not need to for identifying performance of auto-segmentation of objects in images.

Therefore, relating this concern back to the field of radiation oncology, two predicted tumor contours with the same DSC will both be equally encouraged even if one contour disagrees closer to the OAR, resulting in a potentially increased probability of toxicity to the OAR given the segmentation is accepted for radiation dose plan creation. Further, only one prior work to the author's knowledge has attempted weighting the DSC for radiotherapy applications, however the dose was considered which necessitates unmanageable computational resources at this time [3]. Hence, in this work we introduce the concept and formulation of the OAR-Weighted DSC (OAR-DSC) which accounts for spatial location of disagreements between ground truth and auto-contouring generated contours in relation to the surrounding OARs in 2D.

## 2 Materials and Methods

To account for surrounding OARs, the DSC can be rearranged and calculated as follows in Figure 2.

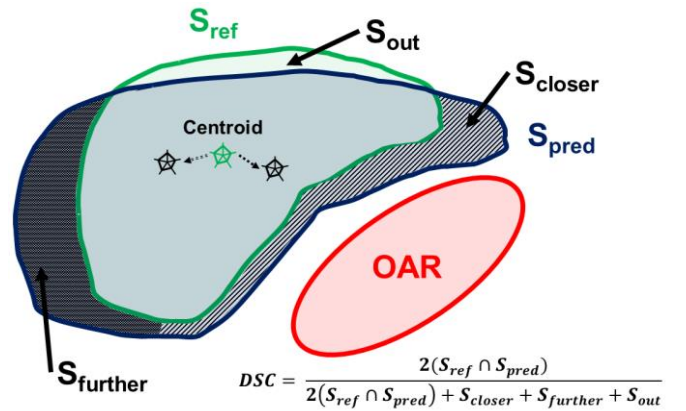

*Figure 2: Rearrangement of the formulation of the DSC between the reference contour, $S_{ref}$, and the predicted auto-contour, $S_{pred}$, to account for nearby OARs. This*

*expression can be weighted to achieve the OAR-DSC as will be shown below.*

This formulation is further described where $S_{ref}$ is the reference ground truth segmentation, $S_{pred}$ is the predicted auto-contouring, $S_{closer}$ is the joint region of voxels not in $S_{ref}$ whose inclusion brings the centroid of $S_{ref}$ closer to the OAR, $S_{further}$ is the joint region of voxels not in $S_{ref}$ whose inclusion brings the centroid of $S_{ref}$ further from the OAR, and $S_{out}$ is the remaining region in the reference ground truth segmentation not covered by the predicted auto-contouring. This also applies to potential voxels inside $S_{ref}$ which overlap with the OAR; however, these will receive the maximum penalty weighting as will be shown below in the description of the OAR-DSC.

To account for predicted contour proximity to the surrounding OAR, a weighting factor, $w_{OAR}$, was applied to $S_{closer}$ as a function of its distance to the OAR to artificially decrease the DSC in cases of predicted contours closer to the OAR than the reference contour. Similarly, a weighting factor, $w_{out}$, was applied to $S_{out}$ as a function of its distance respective to the predicted contour and reference contour to artificially decrease the DSC in cases of target under-coverage. These weighting factors cannot be greater than 1 and no weighting factor is applied to $S_{further}$ to counteract the effects of $S_{closer}$ since this could lead to artificially inflated DSC which may be misleading.

The weighting factor is determined with a simple parameterized exponential decay and scaled if in the presence of multiple OARs based on the maximum distance from the reference contour and the nearest distance to the OAR, $d$. A detailed derivation for a given OAR (i.e., $OAR_i$) is given:

$$OAR\text{-}DSC = \frac{2(S_{ref} \cap S_{pred})}{2(S_{ref} \cap S_{pred}) + \frac{S_{closer}}{w_{OAR^i}} + S_{further} + \frac{S_{out}}{w_{out^i}}}$$

where:

$$w_{OAR^i} = \frac{1}{n_{S_{closer}}} \sum_{j=1}^{n_{S_{closer}}} e^{\left[\alpha \frac{d_{max,OAR^i}}{d_{max,OAR}} \left(\frac{d^j_{pred}}{d^j_{pred}+d^j_{ref}}\right) - 1\right]}$$

and:

$$w_{out^i} = \frac{1}{n_{S_{out}}} \sum_{j=1}^{n_{S_{out}}} e^{\left[\beta \frac{d_{max,out^i}}{d_{max,out}} \left(\frac{d^j_{pred}}{d^j_{pred}+d^j_{ref}}\right) - 1\right]}$$

In this representation, an iteration across all the voxels in $S_{closer}$ is done and at each step, $d_{max,OAR^i}$ is given as the maximum of all the nearest distances from the reference contour to the given $OAR_i$, $d_{max,OAR}$ is given as the maximum of all the nearest distances from the reference contour to the OAR across all considered OARs, $d^j_{pred}$ is the nearest distance from the current voxel to $S_{pred}$, and $d^j_{ref}$ is the nearest distance from the current voxel to $S_{ref}$. The α parameter is representative of the radiation sensitivity of the OAR, therefore increasing values of α will result in decreasing OAR-DSC for the same case. On the other hand, the β parameter is representative of the potential for target spread due to under-irradiation, therefore increasing values of β will result in decreasing OAR-DSC for the same case. It is important to note that when α = 0, then $w_{OAR^i} = 1$ and when β = 0, then $w_{out^i} = 1$, which implies that the OAR-DSC is simply equivalent to the DSC. The OAR-DSC can be computed for each OAR and averaged together to provide a comprehensive score to account for all OARs. A graphical summary of the steps required to calculate the OAR-DSC can be seen below in Figure 3.

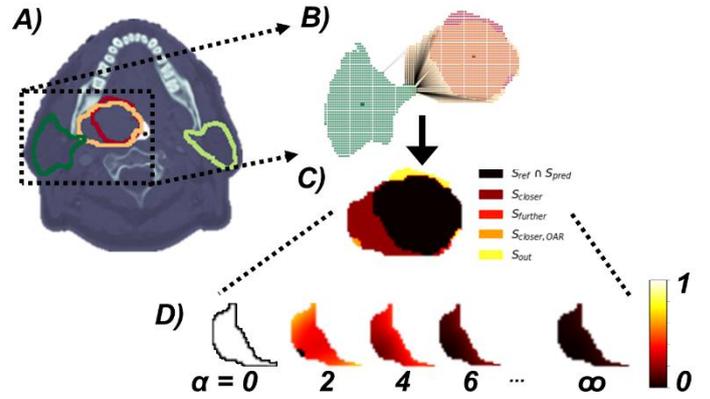

*Figure 3: Detailed graphical summary of the derived maps for the computation of the OAR-DSC of the head and neck region. A) desired target and OAR contours are shown on an anatomical Computed Tomography (CT) image, B) the contours and their voxels are extracted and distances are computed to the OAR, C) voxels are classified into the relevant categories for the OAR-DSC, and D) weighted maps are computed for the voxels inside of $S_{closer}$. Note, the map for α = 0 is all 1 which is white and is represented with a black contour.*

Example cases of the OAR-DSC applied in different anatomical sites was done by utilizing the open-source Contouring Collaborative for Consensus in Radiation Oncology (C3RO) segmentations for the head and neck, gynecological, and gastrointestinal sites [4]. Ground truth reference contours and OAR contours were given by the Simultaneous Truth and Performance Level Estimation (STAPLE) expert segmentations while simulated auto-contouring contours were given by select non-expert segmentation cases. The time required to compute the OAR-DSC compared to the DSC is negligible (~5 seconds for two OARs and six α values) due to the vectorization of the operations. The code for calculating the OAR-DSC given contours in Neuroimaging Informatics Technology

Initiative (NIfTI) format with some examples from the C3RO segmentations is freely available at: https://github.com/Lucas-Mc/OAR-weighted_Dice-Score.

## 3 Results

Applying the OAR-DSC to a few sample cases from C3RO illustrates the following results.

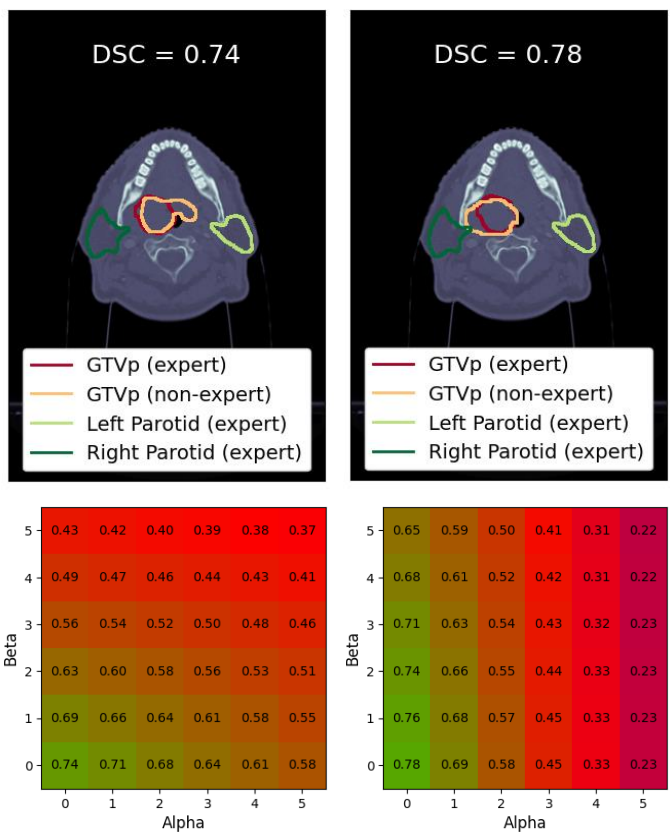

*Figure 4: Comparison of two head and neck cancer cases with similar DSC for the GTVp, yet different proximity to the OARs (left and right parotid). The cumulative OAR-DSC was calculated for each case at different α and β levels, demonstrating increased consideration for the OARs in the right case.*

Figure 4 show a comparison for a head and neck cancer case of two contours with similar DSC scores, yet very different implications for the treatment planning, delivery, and control of toxicity due to the proximity to the OARs (left and right parotid). In fact, the predicted target contour for the primary Gross Tumor Volume (GTVp) which invades the left parotid OAR is rated with a higher unweighted DSC (α = 0), thus being preferred in a deep learning training session. Therefore, the OAR-DSC was calculated for each at different α levels, demonstrating increased consideration for the OARs in the right case as its OAR-DSC falls below the left case with increasing α values. The left case has greater under-coverage of the target and thus has a decreasing OAR-DSC at higher β levels.

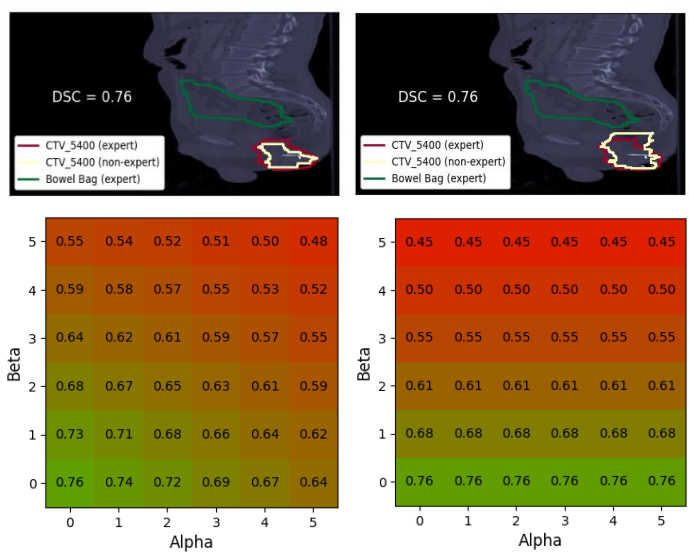

*Figure 5: Examples of OAR-DSC as it applies to the gastrointestinal system.*

Figure 5 shows a sample case for the gastrointestinal region with the Clinical Target Volume to receive 5400 cGy (CTV_5400) set as the target volume and bowel bag set as the OAR. Although both the left and right cases have contours with a DSC of 0.76, the contour in the right case extends closer to the OAR and thus has a decreasing OAR-DSC at higher α levels. The left case keeps the same OAR-DSC throughout since it is entirely contained within the reference contour. However, the contour in the left case has greater under-coverage of the target and thus has a decreasing OAR-DSC at higher β levels.

## 4 Discussion

Since the OAR-DSC inherently includes the DSC at α and β values of 0, this metric has demonstrable translatability to current clinical practice due to simple adjustments of α moving from traditional DSC to an OAR-weighted DSC. However, consideration should be given to standardize the α values and correlate them with validated radiation sensitivity outcomes, such as those provided by QUANTEC including mean and max dose constraints which could be translated to scalable α values [5]. The value of β may be more difficult to find consensus since it depends on the specific application and how concerning under-coverage of the target is.

The limitation of this approach is that care should be taken to account for all desired OARs during the determination of the OAR-DSC, otherwise the auto-contouring algorithm will be unaware of their presence and possibly suggest contours near the OARs. However, this is analogous to the current practice where no considerations are taken for OARs and even more considerate since at least some weighting is being done. Further, this formulation does not specify requirements for utilizing the Gross Tumor Volume

(GTV), Clinical Target Volume (CTV), Planning Target Volume (PTV), or Planning organ at Risk Volume (PRV), however these could be included as the boundaries for the segmentations and used in the same way. This practice, however, should be standardized to ensure comparable OAR-DSC across anatomical sites and centers. It is possible to extend the formulations introduced here to add separate weightings to GTV and CTV or OAR and PRV if equal concern is not desired for the margins as compared to the definite volumes. The formulation also currently uses relative distances between the voxels and OARs to determine radiation sensitivity fraction, but future work could apply absolute distance-based weighting instead based on radiobiological concerns and setup uncertainty (i.e., less than 2 mm should receive the maximum loss weighting). A final limitation of the OAR-DSC is that the OAR segmentations would have to be generated first in a fully automated contouring workflow so that they can be considered when generating the target contours.

Future work includes investigating alternative metrics which could be modified to fit in this framework, such as the surface DSC, and how the OAR-DSC can be expanded to enable more computationally efficient workflows [6, 7]. Additionally, a 3D implementation of the OAR-DSC would enable consideration of all surrounding OARs instead of just those which are in-plane.

## 5 Conclusion

We have introduced the OAR-DSC, a novel method to account for location-specific deviations between an auto-contouring contour and reference contour in relation to surrounding OARs in 2D. Thereby, cases where two proposed auto-contours have similar DSC, we have shown that the OAR-DSC differs between the two especially in cases where one contour expands closer to the surrounding OARs. Additionally, penalizing under-coverage of the target is also addressed due to the nature of the splitting formalism presented for the OAR-DSC vs. the traditional DSC. This metric can be used by deep learning auto-contouring algorithms as a method to add a radiation therapy specific loss function to the generated contours, progressing on its current disregard for the critical influence of these differences on the final radiation dose plan and its risks to the probability of patient toxicity.

## Acknowledgments

Research support was provided by the NIH/NCI Cancer Center Support Grant (CCSG) Pilot Research Program Award from the UT MD Anderson CCSG Image-Driven/Biologically-Informed Therapy Program (IDBT) (PIs Brock/Fuller, P30CA016672-43S1). Lucas McCullum is support by an NIH Diversity Supplement (3R01CA257814-02S2). Kareem A. Wahid is supported by the NCI NRSA Image Guided Cancer Therapy Training Program (T32CA261856). Barbara Marquez is supported by the American Legion Auxiliary Fellowships in Cancer Research and funded by UTHealth Innovation for Cancer Prevention Research Training Program Pre-doctoral Fellowship (Cancer Prevention and Research Institute of Texas grant #RP210042).